\documentclass[aps,prl,twocolumn,groupedaddress,showpacs]{revtex4}

\usepackage{graphicx}
\usepackage{dcolumn}
\usepackage{bm}

\begin{document}

\title{Extraordinary sound screening in perforated plates}

\author{H\'{e}ctor Estrada}
\author{Pilar Candelas}
\author{Antonio Uris}
\author{Francisco Belmar}
\author{Francisco Meseguer}
\email[To whom correspondence should be addressed. ]{fmese@fis.upv.es}
\affiliation{Centro de Tecnolog\'ias F\'isicas, Unidad Asociada ICMM- CSIC/UPV, Universidad Polit\'ecnica de Valencia, Av. de los Naranjos s/n. 46022 Valencia, Spain}
\author{F.J. \surname{Garc\'{i}a} de Abajo}
\affiliation{Instituto de \'Optica - CSIC Serrano 121, 28006 Madrid, Spain}

\date{\today}

\begin{abstract}
We report extraordinary effects in the transmission of sound through periodically perforated plates, supported by both measurements and theory. In agreement with recent observations in slit arrays [M. H. Lu et al. Phys. Rev. Lett. \textbf{99}, 174301 (2007)], nearly full transmission is observed at certain resonant frequencies, pointing out at similarities of the acoustic phenomena and their optical counterpart. However, acoustic screening well beyond that predicted by the mass law is achieved over a wide range of wavelengths in the vicinity of the period of the array, resulting in fundamentally unique behavior of the sound as compared to light. The randomness of the hole distribution and the impedance contrast between the fluid and the solid plate are found to play a crucial role.
\end{abstract}
\maketitle
\begin{figure}
 \includegraphics{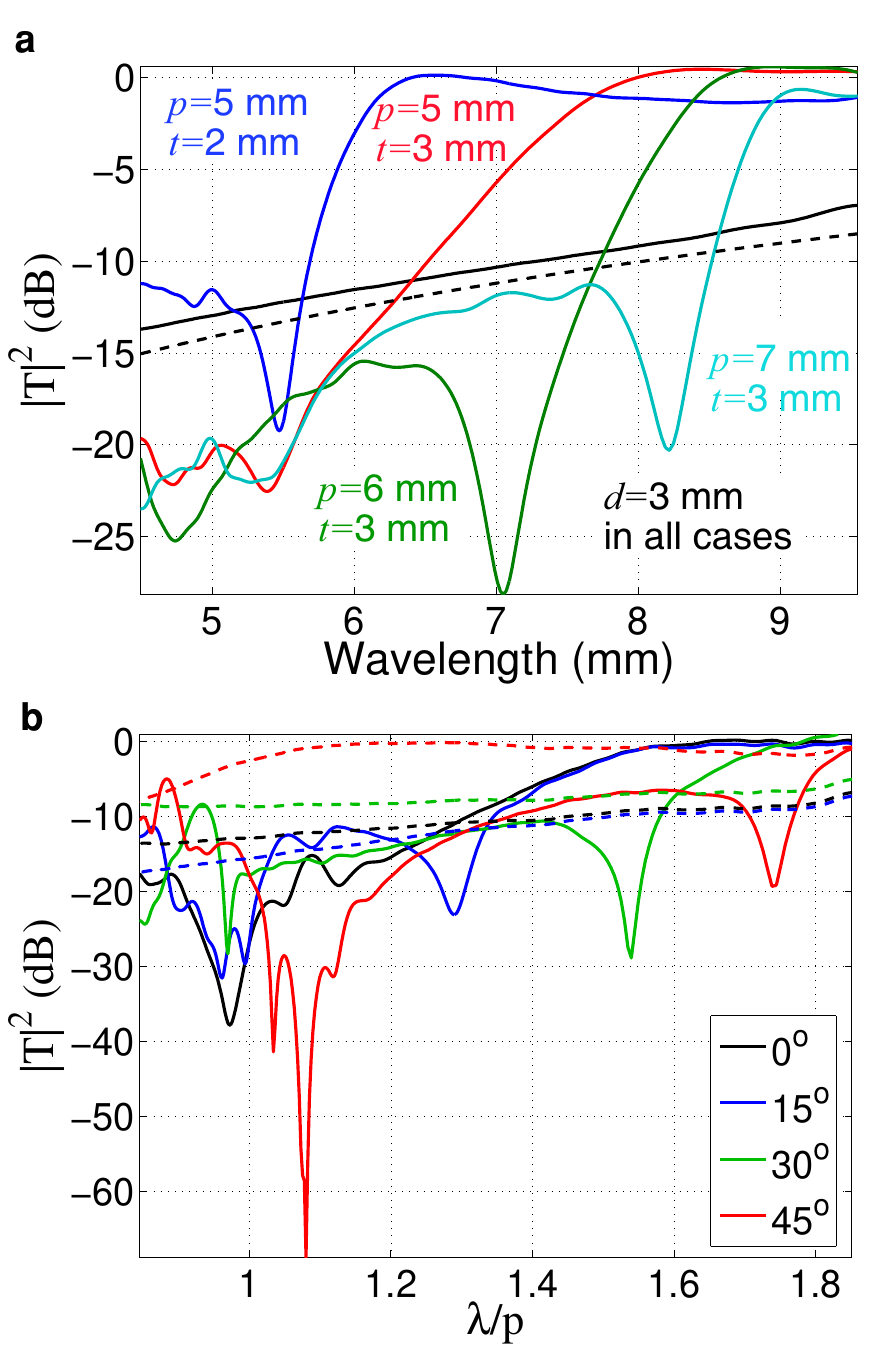}%
 \caption{\label{uno} (a) Measured transmission spectra of Al plates for different geometrical parameters, as indicated by labels. For comparison, we show the transmission of a 3-mm thick plate without holes (black solid curve), along with the prediction of the mass law for that plate (black dashed curve). (b) Transmission spectra of the perforated plate with $d=3$ mm, $p=5$ mm, and $t=3$ mm (solid lines) and the non perforated plate with $t=3$ mm (dashed lines) at different angles of incidence.}
 \end{figure}
\begin{figure}
 \includegraphics[width= \columnwidth]{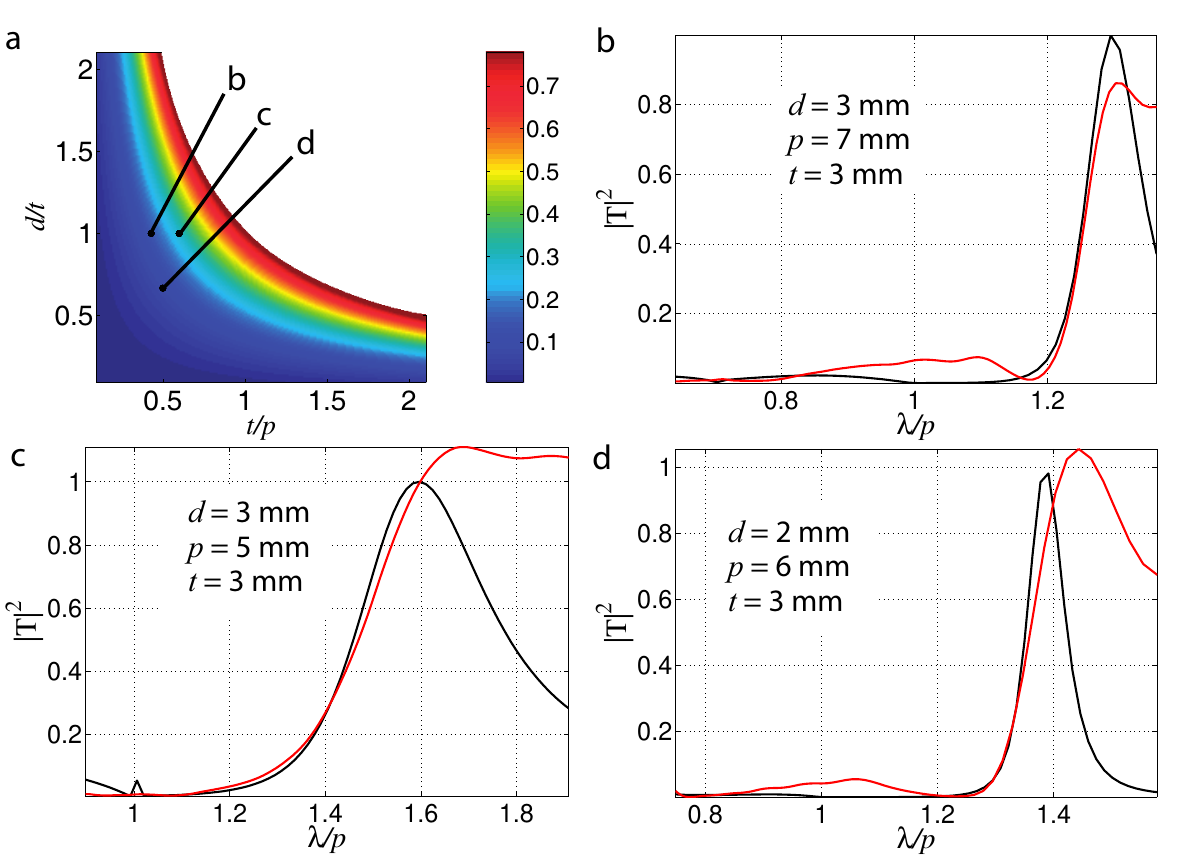}%
 \caption{\label{dos}(a) Filling fraction of perforated plates as a function of t/p and d/t. The dots are the coordinates of measured samples. (b), (c) and (d) Measured (red curves) and calculated (black curves) transmission of arrays with different geometrical parameters (see insets). The measurements are for $37\times14$ holes and the theory corresponds to infinite arrays \cite{messfehler}.}
 \end{figure}

Wave phenomena manifest themselves through different physical realizations \cite{brillouin1953}, ranging from the mechanical nature of sound to the electromagnetic origin of light.  In particular, the enhanced optical transmission observed in metallic membranes pierced by subwavelength hole arrays \cite{ebbenesen98} has prompted interest in areas as diverse as quantum optics \cite{altewischer2002} and negative refraction \cite{ruan2006}.
In the case of acoustic waves, full transmission through subwavelength hole arrays was firstly predicted in \cite{zhou2007} and confirmed experimentally for 1D case in \cite{lu2007}. Similar to light transmission through holes, which is boosted when they are arranged periodically \cite{ebbenesen98}, plates can be made nearly transparent to sound at certain frequencies if they are pierced by a periodic array of apertures. Like in its optical counterpart, this extraordinary acoustic phenomenon occurs for openings much narrower than the wavelength. But in contrast to light, (a) small holes drilled in hard materials can support at least one guided mode, regardless how narrow they are (provided the hole radius remains larger than the viscous skin depth of the fluid), and (b) sound penetrates into the solid depending on the impedance contrast between fluid and plate, making sound unique and giving rise to colorful behavior of perforated plates. We have measured sound transmission in perforated plates immersed in water at ultrasonic frequencies using a transducer to generate a pulse that is normally incident on a plate, transmitted through the sample plate, and detected by another transducer on the far side of the sample. We use a couple of transmitter/receiver ultrasonic Imasonic immersion transducers with 32 mm in active diameter, -6 dB bandwidth between 169-330 kHz (corresponding to wavelengths between 4.5 mm and 8.8 mm in water), and with a far-field distance of 42 mm. A pulser/receiver generator (Panametrics model 5077PR) produces a pulse which is applied to the emitter transducer to launch the signal through the inspected plate. The signal is detected by the receiving transducer, acquired by the pulser/receiver, post amplified, and finally digitized by a digital PC oscilloscope (Picoscope model 3324). Each measure consist in the average over 256 pulses to increase the signal-to-noise ratio. The plates are 200 mm wide, 350 mm long, and clamped during the measurements. Each transducer is located at a distance of 90 mm from the plate and aligned for normal incidence. The transmission spectrum is then obtained from the power spectrum of the signal normalized to the reference signal measured without the plate. Holes were mechanically drilled to form either periodic square arrays or disordered arrays, in plates of PMMA, aluminium and brass. Plates of different thickness $t$, hole diameter $d$, and period of the array $p$ have been examined. Figure \ref{uno}(a) shows transmission spectra for various periodic hole arrays drilled in aluminium plates. Two common distinct features can be observed. First, the transmission is very low at water wavelengths close to the period, a manifestation of Wood anomalies similar to those observed in optical grating \cite{wood1902}, first explained by Lord Rayleigh \cite{rayleigh1907} and ultimately related to the piling up of inter-hole interaction under that condition. Perforated plates can thus shield sound much more effectively than uniform plates. This effect, which is dramatic in the measured $d = 3$ mm, $p = 6$ mm, $t = 3$ mm plate at a wavelength of 7 mm (Fig. \ref{uno}(a)), violates the mass law, shown as a black dashed curve in Fig. \ref{uno}(a) and stating that more massive walls produce more efficient soundproofing \cite{taschenbuch}. This effect is observed not only at normal incidence, but also for different tilted angles as shown in Fig. \ref{uno}(b)
This result can have enormous impact in soundproofing of machines that require efficient cooling, with the latter facilitated by the presence of the holes and achieving around 15 dB more attenuation than a panel without holes. As a second feature, plates become completely transparent to sound for a broad range of wavelengths above the period, exhibiting a maximum of transmission at a wavelength that depends both on the filling fraction occupied by holes and on the ratio between plate thickness and array period. In some cases the transmission spectrum shows step like behavior typical from the low band pass filters. It can have important applications to ultrasonic non destructive testing and to invisible materials at ultrasonic frequencies, as it happens in the optical case \cite{alu2005}. We can gain further insight into the mechanisms of sound transmission in these holey arrays through comparison with model calculations in the hard-solid limit, that is, when an infinite impedance contrast is assumed for the plate/water interface, which effectively translates into a description of sound via a scalar pressure $\phi$ , subject to the wave equation in the water medium and vanishing normal-derivative boundary condition
\begin{equation}
 (\nabla^2 + k^2)\phi  = 0,\ \partial_n \phi=0 \,,\label{helm}
\end{equation}
where $k = 2\pi/\lambda$ is the wavevector and $\lambda$ is the water wavelength. In this limit, the pressure field does not penetrate into the plate material because the plate/water interface is completely rigid, so that the gradient of pressure must be zero along the interface normal. We rigorously solve Eq. (\ref{helm}) using an extension of Takakura's method for light in a single slit (ref. \cite{takakura2001}). Namely, (i) we expand the pressure in terms of guided modes inside the circular hole cavities using Bessel functions with zero derivative at the wall of the holes. (ii) A Rayleigh plane-wave expansion is used on either side of the film. (iii) Inside and outside solutions are matched at the film surfaces to satisfy the vanishing of the normal derivative at the solid surface and the continuity of both the field and its derivative at the hole openings. This results in a set of linear equations involving the expansion coefficients inside and outside the film. (iv) The latter coefficients are expressed in terms of the former, thus reducing the system to a linear set of equations involving only hole-cavity-mode coefficients for a representative hole, which we solve for a finite number of low-order modes. Convergence is achieved for $\sim 20$ guided modes in the cases explored throughout this work. (v) The transmission is finally obtained from the coefficients of the noted expansion. This procedure leads to similar analytical results as in ref. \cite{zhou2007} when only one cavity mode is selected (the propagating, cutoff-free guided mode). We have performed numerically-converged calculations, which are required for the wide set of geometrical parameters $t/p$ and $d/t$ actually tested, as shown by the dots of Fig. \ref{dos}(a), in which the contour plot represents the filling fraction of the holes in the array. Figures \ref{dos}(b)-(d) show some measurements as compared to calculated results. No other approximation beyond the hard-solid limit has been made in our theory (black solid curves), which agrees well with experiment (red curves) and predicts a transmission maximum of 100\% at wavelengths above the period, immediately flanked by a minimum of vanishing transmission on the lower wavelength side. This behavior is reminiscent of Fano resonances produced by coupling of a discrete state to a continuum \cite{fano1961}. Actually, similar Fano resonances have been identified in optical transmission resonances through perforated arrays \cite{genet2003}. In the optical case, the discrete resonance has been assigned to lattice singularities originating in Wood anomalies \cite{abajo2005E,abajo2007,liu2008n}. However, the situation is more complicated in sound.

Indeed, the condition of vanishing normal gradient pressure in the hard-solid limit has interesting implications for the performance of the holes, and in particular, a propagation mode is found confined inside each hole, with uniform pressure along any plane parallel to the plate, acting as an acoustic mass. This mode exists for arbitrarily-large sound wavelength compared to the size of the openings, similar to the TEM optical mode in metallic slits and annular holes \cite{jackson1999,hou2007}. Fabry-P\'erot (FP) resonances in each hole regarded as a cavity are then set up by reflection of this mode at the entry and exit sides of the plate, giving rise to resonant coupling of sound that produces enhanced transmission at specific wavelengths. A similar phenomenon has been observed in transmission through narrow slits \cite{takakura2001,yang2002}. Therefore, and in contrast to light, even a single hole produces transmission resonances of sound. In a hole array, the position of the resonance is dictated by the interplay between the noted FP resonances and the interaction among holes. And similar to its optical counterpart, sound transmission displays Fano profiles as well because it also involves coupling of a discrete resonant state (the hybrid of FP and lattice resonances) with the continuum of sound propagating in the surrounding medium (i.e., water in our case). Although both sound and electron wavefunctions in quantum mechanics respond to the same wave equation in homogeneous regions (e.g., Schrodinger's equation in the latter), the boundary conditions for electron waves involve the vanishing of the field rather than its derivative at the interface with infinite-potential regions (the equivalent of our hard-solid material in sound), thus precluding the existence of guided modes in narrow hole cavities. These two types of waves are thus prototypical examples of transmission of scalar waves through hole arrays in opaque films, showing extreme boundary conditions.

\begin{figure}
 \includegraphics[width= \columnwidth]{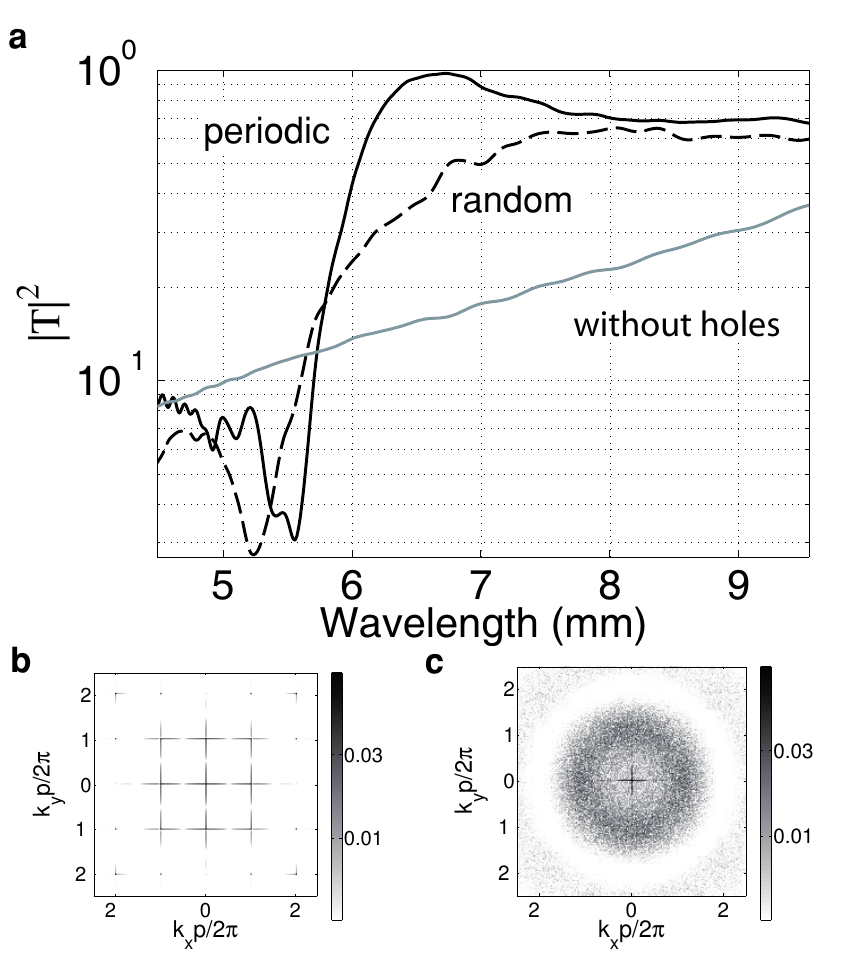}%
 \caption{\label{tres}(a) Transmittance spectra of 2-mm thick aluminium plates pierced by $39\times39$ holes of 3 mm in diameter, distributed periodically (black line, for a period p =5 mm) and randomly (red line). The transmission without holes is shown for comparison (blue line). (b) and (c) 2D Fourier transforms (contour plots in log scale) of the periodic and random arrays, respectively, showing hot spots of the former and a broad distribution of the latter in reciprocal space.}
\end{figure}
\begin{figure}
 \includegraphics[width = 0.8\columnwidth]{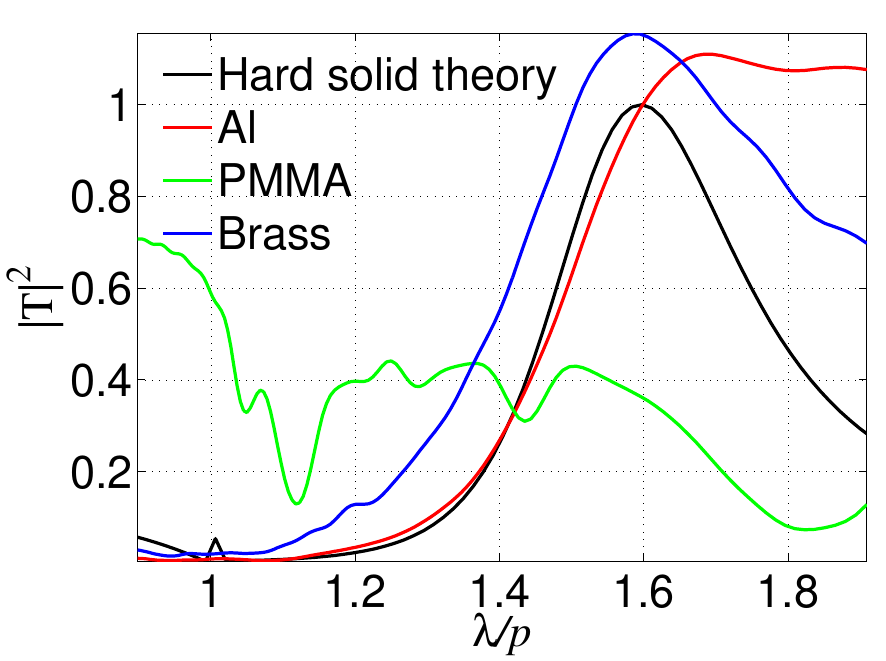}%
 \caption{\label{cuatro}Transmission spectra of the $d = 3$ mm, $p = 5$ mm, $t = 3$ mm perforated plates made of brass (blue curve), aluminium (red curve), and PMMA (green curve), compared to theory in the hard-solid limit (black curve).}
 \end{figure}

Another aspect that makes sound transmission unique with respect to light transmission is the absence of a skin depth effect. In real solids deviating from the hard-solid limit, sound can penetrate and is not attenuated exponentially like light inside a metal. Also the existence of surface waves in the plate (leaky Lamb waves) and interface waves \cite{liu2008} (Scholte waves) between the plate and the fluid can play an important role in the sound propagation, especially if the incidence direction is not normal to the plate. Our sound transmission study would be analogous in that instance to optical transmission through holes in dielectrics of high index of refraction. In other words, conventional materials behave as optical dielectrics for sound, although the equivalent of metallic behavior, characterized by a negative dielectric function, has been also observed in acoustic metamaterials \cite{liu2000}.
The above considerations are supported by further investigating sound transmission in disordered holes arrays, as compared to transmission in ordered arrays (Fig. \ref{tres}). Strong transmission dips and regions of large transmission do occur in both cases, and so does the Fano mechanism discussed above, via FP resonances involving TEM modes. However, the periodicity of ordered arrays facilitates a cooperative effect in the transmission of the holes, since translational invariance guarantees a single resonance for each direction of sound incidence, whereas random arrays have different resonant wavelengths for different hole environments (i.e., the interaction of neighboring holes modifies the reflection coefficients of the hole cavity, which influences in turn the actual wavelength of its FP resonance), thus leading to a transmission spectrum with higher density of features (Fig. \ref{tres}(a)), as well as lower values of the transmission. This behavior is actually supported by sharp versus spread distributions in the Fourier transform of periodic and random hole arrays, respectively (see Fig. \ref{tres}(b),(c), which has been shown to govern the wavelength distribution of spectral features \cite{huang2007}. Deviations from the hard-solid limit are explored in Fig. \ref{cuatro} by comparing transmission spectra of arrays with identical geometry but made of different materials. The highest impedance ratio under consideration is 25 in brass, for which the transmission spectrum follows quite closely the prediction of our hard-solid theory. The second highest impedance ratio is 11.8 for aluminium, which also mimics rather well the theory, except for high values of the transmission at wavelengths beyond the maximum, possibly due to the finite size of the incident wave. Finally, the low impedance contrast of PMMA (only 2.1) results in a completely different spectrum, governed by the interplay between diffraction at the holes and sizeable transmission through the bulk of the plate. Actually, the latter mechanism reveals itself clearly through a 60\% transmission at $\lambda = p$, in contrast to the Wood-anomaly dip when transmission is dominated by sound guiding in the holes.
The large suppression of transmission over relatively wide wavelength regions centered around the lattice period in ordered arrays of holes drilled through metallic plates offers interesting possibilities to shield sound while allowing light to pass through the holes. There are several transmission mechanisms involved in sound passing through drilled plates, which make them more complicated as compared to other types of waves, and which produce interesting interplay phenomena: (1) transmission assisted by cutoff-free waveguide modes of individual holes; (2) interaction among holes in periodic (or aperiodic) arrays; (3) direct transmission through the bulk material, in which the absence of a skin-depth effect allows appreciable contributions in low-impedance-contrast materials, even for relatively large plate thicknesses. Control over these different mechanisms can find potential application to mimic with sound similar effects as those realized in photonic metamaterials, like cloaking \cite{pendry2006,cai2007}, subwavelength imaging \cite{liu2007}, light focusing via the Talbot effect \cite{huang2007}, and resonant wavelength filtering \cite{han2006}.

\begin{acknowledgments}
This work has been partially supported by the Spanish CICyT
(projects  MAT2006-03097; MAT2007-66050, and Consolider CSD2007-00046), Intramural Project  CSIC Nr.:200560F0071. H. Estrada acknowledges CSIC-JAE scholarship.\\
\end{acknowledgments}


\begin{thebibliography}{27}
\bibitem[{mes()}]{messfehler}
\bibinfo{note}{Normalization leads to transmission values slightly above 100\%
  in some cases, which we attribute to the finite size of the incident wave so
  that the wave front generated by the transducer is not perfectly plane.
  Similarly, finite size effects can appear in the measurements if the number
  of holes is small. Actually, these effects can be minimized using large
  arrays as those used in Fig. \ref{tres}(a) ($39\times39$).}
\end{thebibliography}
\end{document}